\documentclass[journal,twoside,a4paper]{IEEEtran}
\usepackage{amsmath,amsfonts}
\usepackage{algorithmic}
\usepackage{algorithm}
\usepackage{array}
\usepackage[caption=false,font=normalsize,labelfont=sf,textfont=sf]{subfig}
\usepackage{textcomp}
\usepackage{stfloats}
\usepackage{url}
\usepackage{verbatim}
\usepackage{graphicx}
\usepackage{cite}
\usepackage{hyperref}
\usepackage[capitalise,nameinlink]{cleveref}
\crefname{equation}{}{}
\Crefname{figure}{Fig.}{Figs.}
\usepackage{latexml}
\usepackage{eso-pic} 
\begin{document}
	\def\copyrightdoi{10.1109/TASC.2025.3526746}
	\def\copyrightref{IEEE Transaction on Applied Superconductivity}
	\def\copyrightyear{2025}
	\iflatexml
	This article has been accepted for publication in \textsc{\copyrightref}. This is the author's version which has not been fully edited and content may change prior to final publication. Citation information: DOI \href{http://dx.doi.org/\copyrightdoi}{\copyrightdoi}.
	
	\textcopyright~\copyrightyear~IEEE. Personal use of this material is permitted.  Permission from IEEE must be obtained for all other uses, in any current or future media, including reprinting/republishing this material for advertising or promotional purposes, creating new collective works, for resale or redistribution to servers or lists, or reuse of any copyrighted component of this work in other works.
	\else
		\AddToShipoutPictureBG{%
		\AtPageLowerLeft{%
			\hspace*{\dimexpr0.5\paperwidth\relax}
			\makebox(0,3cm)[c]{\parbox{0.8\paperwidth}{%
					\footnotesize \textcopyright~\copyrightyear~IEEE. Personal use of this material is permitted.  Permission from IEEE must be obtained for all other uses, in any current or future media, including reprinting/republishing this material for advertising or promotional purposes, creating new collective works, for resale or redistribution to servers or lists, or reuse of any copyrighted component of this work in other works.}}%
		}%
		\AtPageUpperLeft{
			\hspace*{\dimexpr0.5\paperwidth\relax}
			\makebox(0,-1cm)[c]{\parbox{0.8\paperwidth}{%
					\centering \footnotesize This article has been accepted for publication in \textsc{\copyrightref}. This is the author's version which has not been fully edited and content may change prior to final publication. Citation information: DOI \href{http://dx.doi.org/\copyrightdoi}{\copyrightdoi}.}}%
		}%
	}%
	\fi
	\title{Obstacles to two-line Josephson traveling wave parametric amplifier}
	\author{Victor K. Kornev, Alena N. Nikolaeva, Nikolay V. Kolotinskiy%
		\thanks{Manuscript received July 25, 2024; revised November 05, 2024 and January 02, 2025.}%
		\thanks{This research was funded by Russian Science Foundation (RSCF) grant no.~19-72-10016-P.}
		\thanks{Victor K. Kornev, Alena N. Nikolaeva and Nikolay V. Kolotinskiy are with Department of Physics, Lomonosov Moscow State University, 119991, Moscow, Russia. Any correspondence should be addressed to Nikolay V. Kolotinskiy (\protect\href{mailto:kolotinskij@physics.msu.ru}{kolotinskij@physics.msu.ru}).}}
	
	\markboth{S\MakeLowercase{ubmitted to} IEEE Transaction on Applied Superconductivity}%
	{Kornev \MakeLowercase{\textit{et al.}}: Two-line Josephson traveling wave parametric amplifier}
	
	\maketitle
	
	\begin{abstract}
		Obstacles to the two-line Josephson traveling wave parametric amplifier
		(JTWPA) designs, aimed at increasing the allowed pump wave energy and,
		hence, the gain growth through the combination of linear and nonlinear
		lines as pump and signal lines is discussed. This analysis takes into
		account both the mutual coupling and the discreteness of the artificial
		waveguide lines. Significant restrictions arise from the cyclical
		transfer of the traveling wave's energy from one line to the other one
		with phase step by $\pi$ every beat cycle. Moreover, the other
		restrictions result from the phase mismatch between the pump, signal and
		idler waves due to both the coupling and the discreteness of the
		artificial lines.
	\end{abstract}

		\section{Introduction}
\begin{figure}[t]
		\centering
		\includegraphics[width=7.5cm]{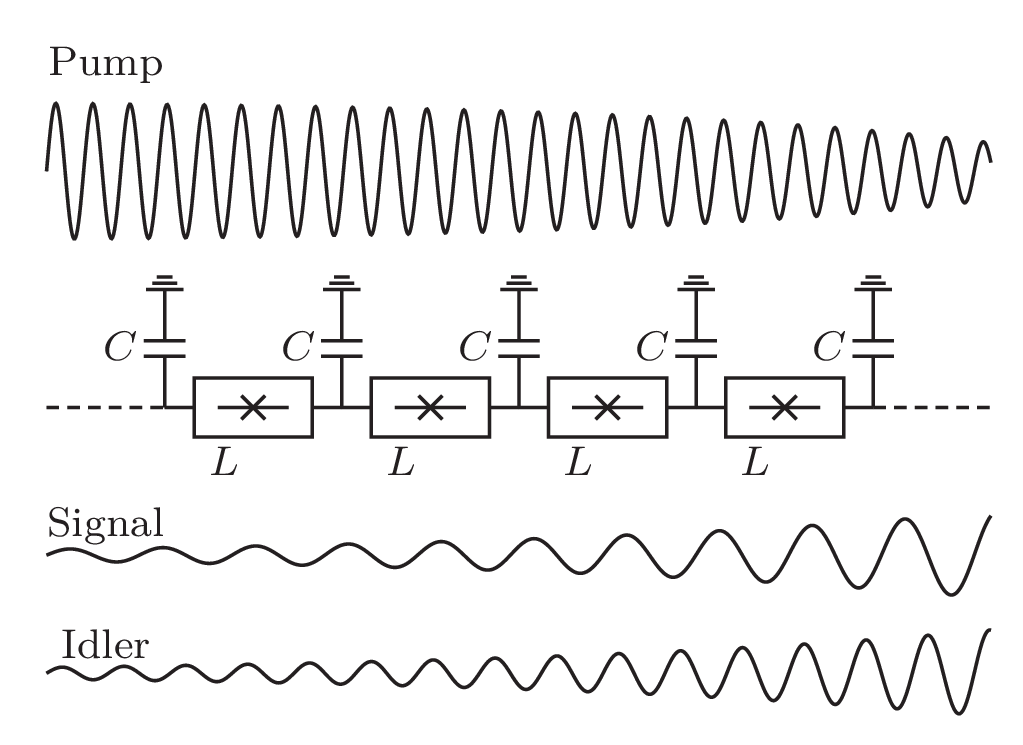}
		\caption{\label{fig:JTWPA_idea}Illustration of a TWPA composed of a chain of Josephson-junction cells with propagating pump, signal, and idler waves.}
	\end{figure}

	\IEEEPARstart{T}{he} parametric amplification mechanism enables extremely
	high-sensitive amplifiers capable of showing noise temperature
	$T_N$ lower than the ambient temperature. A traveling wave design concept of
	the amplifiers allows overcoming the gain-bandwidth trade-off which is
	attributable to the cavity-based parametric amplifiers. Superconductive
	Josephson Traveling Wave Parametric Amplifiers (JTWPAs) are capable of
	working at low and very low temperatures and capable of approaching
	quantum limit sensitivity. Therefore, the amplifiers are considered
	promising readout devices in the fields of precision quantum
	measurements, quantum communications and quantum computing (\textit{e.g.}, see
	\cite{Macklin2015, Haider2019, FadaviRoudsari2023}, and review \cite{Esposito2021}). As shown schematically in \cref{fig:JTWPA_idea}, these amplifiers are based on the use of artificial discrete waveguide lines of LC-cells each composed of a linear
	capacitor $C$ and nonlinear $L$-element. The latter is usually
	either a single Josephson junction or a Superconducting Quantum Interference
	Device (SQUID containing one, two or several Josephson junctions. In the
	general case, the \emph{L}-element can combine both the geometric
	inductance and the strongly nonlinear kinetic inductance of the Josephson
	junctions
	\begin{equation}
		L_{J} = L_{J0}/\cos\left(\varphi\right)
	\end{equation}
	which is attributable to the superconducting current component of the
	used junctions:
	\begin{equation}
		I_{s} = I_{c}\sin\left(\varphi\right), 
	\end{equation}
	where $I_{c}$ and $\varphi$ are the junction critical current and
	phase, respectively;
	$L_{J0} = \Phi_{0}/\left(2\pi I_{c}\right)$ is
	the characteristic value of the junction inductance, and
	$\Phi_{0} = h/(2e)$ is the magnetic flux quantum.
	
	Nonlinear interaction between the pump and signal waves applied
	initially to the line input originates an idler wave. Subsequent
	nonlinear interaction of three waves propagating along the line yields
	in the energy transfer from the high-power pump wave to both the signal
	and idler waves. The relation between frequencies $\omega_p$, $\omega_s$, $\omega_i$ of the pump, signal and idler waves,
	respectively, depends on the exploited mixing mode. Quadratic
	nonlinearity allows the use of a three-wave-mixing mode (3WM), when the
	frequencies obey the relation
	\begin{equation}
		\label{eq:3W}
		\omega_s+\omega_i=\omega_p
	\end{equation}
	and the wave vectors obey the relation
	\begin{equation}
		\label{eq:3Wk}
		k_p=k_s+k_i
	\end{equation}
	needed to achieve the required phase matching of the waves in the 3WM
	mode.
	
	Cubic (Kerr) nonlinearity allows the use of four-wave-mixing mode (4WM)
	when the frequencies obey the following relation:
	\begin{equation}
		\label{eq:4W}
		\omega_s+\omega_i=2\omega_p
	\end{equation}
	and the wave vectors obey the relation
	\begin{equation}
		\label{eq:4Wk}
		2k_{p} = k_{s} + k_{i}
	\end{equation}
	needed to achieve the required phase matching of the waves in the 4WM
	mode.
	
	The achievable gain depends on several factors \cite{Dixon2020, Peng2022, Peng2022b, Remm2023}, among which
	is the pump wave depletion with the energy transfer to the signal and
	idle waves, as well as to the higher pump harmonics and undesired
	intermodulation components \cite{Zorin2016, Zorin2017}. The depletion cannot be balanced	with the starting power of the pump wave since the pump wave amplitude is restrained by the critical-current value of the Josephson junctions used.

	\begin{figure}[t]
		\centering
		\includegraphics[width=7cm]{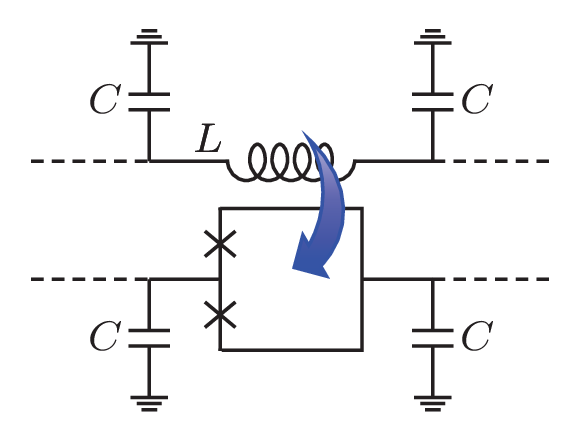}
		\caption{\label{fig:JTWPA}A fragment of the flux driven JTWPA \cite{Zorin2019} composed of two artificial waveguide lines, which are (1) linear LC line and (2) nonlinear Josephson line containing dc SQUIDs as nonlinear $L$-elements. In the design, the pump wave current, flowing in the linear inductances $L$ of the first line, applies magnetic flux to the dc SQUID cells of the second (signal) line. This provides modulation of the signal line cells inductances in a traveling wave pattern.}
	\end{figure}
	
	To overcome the gain limitation, a flux-driven JTWPA has been proposed
	\cite{Zorin2019} although is not still realized. The amplifier is composed of
	two artificial lines, namely of a nonlinear Josephson line and a linear
	LC line meant for propagation of a high-power pump wave. The pump wave
	current, flowing in the inductances L of the linear line, applies
	magnetic flux to the dc SQUID cells of the second (signal) line as shown
	in \cref{fig:JTWPA}. This results in modulation of the cell inductances along the line in a traveling wave pattern \cite{Zorin2019,Nikolaeva2023} as follows:
	\begin{equation}
		L=L_0\left(1+m\cdot\cos\left(\omega_pt-k_px\right)\right),
	\end{equation}
	where $m$ is modulation depth, and $L_0$ is the reference SQUID-cell inductance controlled by the dc magnetic flux which is
	additionally applied. Such a modulation relation allows realizing the
	three-wave-mixing mode \cref{eq:3W}, when the required phase-matching condition
	\cref{eq:3Wk} is fulfilled as well.
	
	The flux driven JTWPA can be considered as a particular case of a more
	general design concept of the two-line JTWPAs, composed of a nonlinear
	Josephson waveguide line and a linear artificial LC waveguide line meant
	for propagation of a high-power pump wave. The linear line is supposed to be used as a pump wave energy source for the nonlinear Josephson line in order to balance (through coupling elements) the pump wave power depletion and hence to support the fixed pump wave amplitude in this waveguide line. However, as a matter of fact, this design concept implies
	realizing a completely nonreciprocal coupling between the two waveguide
	lines. Unfortunately, this cannot be obtained at least with using the
	linear electromagnetic linkage. Therefore, one should take into account
	the inevitable mutual coupling between the two waveguide lines.
	
	For example, practically all the real designs meant to answer to the
	schematic shown in \cref{fig:JTWPA} for the flux-driven JTWPA give concurrently
	two effects. Namely, the pump wave current in the first (linear) line
	concurrently (i) apples magnetic flux to the dc SQUID cells of the
	second line and (ii) induces an electromotive difference along the cell
	as the nonlinear inductance of the second (signal) line. The latter
	results from the inevitable spatial design asymmetry. In fact, when the
	cell inductance $L$ of the first line is coupled magnetically mainly
	with the nearest loop side of the SQUID cell, the pump wave current
	$I_{p} = I_{0}e^{- i\omega_{p}t}$ induces in the loop part the
	electromotive difference $V_{em} = i\omega_{p}MI_{p}$, where $M$ is
	the mutual inductance between this loop part and inductance $L$. This
	electromotive difference produces both the circular ac current and ac
	voltage drop across the SQUID cell (between opposite points of the
	cell):
	\begin{equation}
		\label{eq:VD}
		U=\frac{1}{2}M\frac{dI_p}{dt}=i\omega_p\frac{M}{2}I_p.
	\end{equation}
	
	This expression evidences the mutual coupling between these waveguide
	lines through the effective mutual inductance $M/2$ between cells of
	the lines.
	
	In common with two coupled oscillatory circuits, such coupling between
	the waveguide lines radically alters the whole wave process in the
	system. Given the increased interest in possible JTWPA designs based on
	two artificial lines such as the proposed flux-driven JTWPA, it is
	completely reasonable to consider the wave processes occurring in the
	coupled lines in terms applicable to the JTWPA designs.
	
	In this paper, we analyze the obstacles to the two-line JTWPA designs
	taking into account both the mutual coupling and discreteness of the
	artificial lines used.
	
	\section{Consideration of two coupled linear waveguides}
	
	A comprehensive study of the traveling wave parametric amplifiers based
	on the use of nonlinear waveguide lines must be conducted using
	numerical simulations or by solving a complex set of the
	coupled-mode-equations (\textit{e.g.} see \cite{Dixon2020}). The presence of
	nonlinear reactive elements, which are needed for the parametric
	amplification process, complicates the device dynamics through
	generation of both the wanted and unwanted spectral components. However,
	the device functioning is inherently based on the wave propagation
	process, which can be radically changed by coupling between the
	waveguide lines. Basically, this change can be studied through
	considering two coupled linear waveguide lines as done below.
	
	\Cref{fig:coupledlines} shows equivalent circuits of two
	coupled linear waveguide lines with capacitive (a) and inductive (b)
	couplings realized through either linking capacitance $C_0$ or the mutual inductance $M$, respectively. In the continuum approximation, the telegraph
	equations for the first system can be written as follows:
	
	\begin{figure*}[t!]
		\centering
		\subfloat[]{\includegraphics[width=7.5cm]{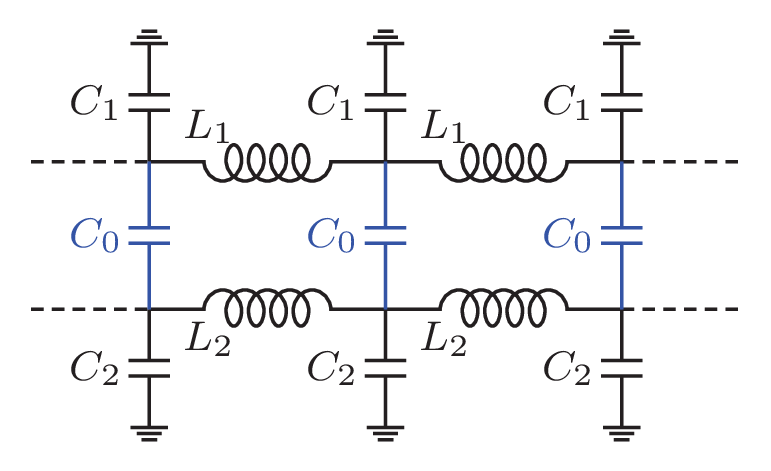}}\qquad
		\subfloat[]{\includegraphics[width=7.5cm]{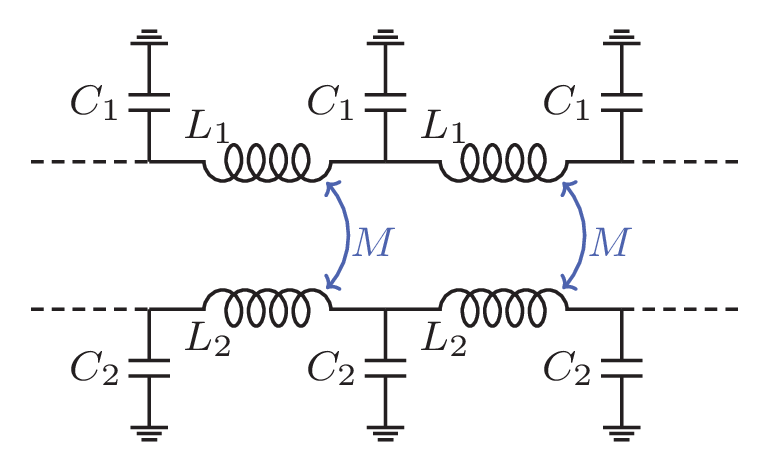}}
		\caption{\label{fig:coupledlines}Equivalent circuits of two coupled linear waveguide lines with capacitive (a) and inductive (b) couplings realized 	through either the linking capacitance $C_0$ or the
			mutual inductance $M$, respectively.}
	\end{figure*}
	
	\begin{align}
		\label{eq:system1b}
		\frac{\partial U_{1}}{\partial x} = - L_{1}\frac{\partial I_{1}}{\partial t},&\qquad
		\frac{\partial I_{1}}{\partial x} = - C_{1}\frac{\partial U_{1}}{\partial t} - C_{0}\frac{\partial{(U}_{1} - U_{2})}{\partial t}, \\ 
		\frac{\partial U_{2}}{\partial x} = - L_{2}\frac{\partial I_{2}}{\partial t},&\qquad
		\frac{\partial I_{2}}{\partial x} = - C_{2}\frac{\partial U_{2}}{\partial t} - C_{0}\frac{\partial{(U}_{2} - U_{1})}{\partial t},
		\label{eq:system1e}
	\end{align}
	where $U_{1,2}$ and $I_{1,2}$ are the voltage and the current in the
	first and second lines, respectively. These equations yield in the
	following set of two coupled wave equations:
	
	\begin{align}
		\label{eq:system2b}
		v_{01}^{2}\frac{\partial^{2}U_{1}}{\partial x^{2}} &= \frac{\partial^{2}U_{1}}{\partial t^{2}} - \alpha_{1}\frac{\partial^{2}U_{2}}{\partial t^{2}}, \\
		v_{02}^{2}\frac{\partial^{2}U_{2}}{\partial x^{2}} &= \frac{\partial^{2}U_{2}}{\partial t^{2}} - \alpha_{2}\frac{\partial^{2}U_{1}}{\partial t^{2}},
		\label{eq:system2e}
	\end{align}
	where $v_{01}^{2} = \left( L_{1}C_{1} \right)^{- 1}$,
	$v_{02}^{2} = \left( L_{2}C_{2} \right)^{- 1}$ are the squared phase velocities in the separate (uncoupled) lines,
	$\alpha_{1} = C_{0}/(C_{1} + C_{0}) \approx C_{0}/C_{1}$,
	$\alpha_{2} = C_{0}/(C_{2} + C_{0})\approx C_{0}/C_{2}$ are the
	capacitive coupling factors of the lines. In case of inductive coupling,
	telegraph equations are as follows:
	\begin{align}
		\frac{\partial U_{1}}{\partial x} = - L_{1}\frac{\partial I_{1}}{\partial t} - M\frac{\partial I_{2}}{\partial t},&\qquad
		\frac{\partial I_{1}}{\partial x} = - C_{1}\frac{\partial U_{1}}{\partial t}, \\
		\frac{\partial U_{2}}{\partial x} = - L_{2}\frac{\partial I_{2}}{\partial t} - M\frac{\partial I_{1}}{\partial t},&\qquad
		\frac{\partial I_{2}}{\partial x} = - C_{2}\frac{\partial U_{2}}{\partial t},
	\end{align}
	and yield in the similar set of two coupled wave equations:
	\begin{align}
		\label{eq:system4b}
		v_{01}^{2}\frac{\partial^{2}I_{1}}{\partial x^{2}} &= \frac{\partial^{2}I_{1}}{\partial t^{2}} - \eta_{1}\frac{\partial^{2}I_{2}}{\partial t^{2}}, \\
		v_{02}^{2}\frac{\partial^{2}I_{2}}{\partial x^{2}} &= \frac{\partial^{2}I_{2}}{\partial t^{2}} - \eta_{2}\frac{\partial^{2}I_{1}}{\partial t^{2}}, \label{eq:system4e}
	\end{align}
	where $v_{01}^{2} = \left( L_{1}C_{1} \right)^{- 1}$,
	$v_{02}^{2} = \left( L_{2}C_{2} \right)^{- 1}$ are the squared phase velocities in the separate (uncoupled) lines, $\eta_{1} = M/L_{1}$, $\eta_{2} = M/L_{2}$
	are the inductive coupling factors of the lines.
	
	Due to the complete similarity between the wave equations sets \labelcref{eq:system2b}, \labelcref{eq:system2e} and \labelcref{eq:system4b}, \labelcref{eq:system4e}, one can consider the only first equations set and look for its solution as
	\begin{equation}
		U_{1} = A\cos\left(\omega t - kx\right),\quad U_{2} = B\cos\left(\omega t - kx\right).
	\end{equation}
	
	In this case, \cref{eq:system2b,eq:system2e} gives the following system of two linear equations for the wave amplitudes:
	
	\begin{align}	
		\label{eq:system5b}
		\left( k^{2}v_{01}^{2} - \omega^{2} \right)A + \alpha_{1}\omega^{2}B &= 0,\\	
		\alpha_{2}\omega^{2}A + \left( k^{2}v_{02}^{2} - \omega^{2} \right)B &= 0.
		\label{eq:system5e}
	\end{align}
	
	Equating determinant of the equations set to zero, one comes to the
	following master equation
	\begin{equation}
		\label{eq:meq}
		\left( k^{2}v_{01}^{2}\text{-}\omega^{2} \right)\left( k^{2}v_{02}^{2}\text{-}\omega^{2} \right) - \alpha_{1}\alpha_{2}\omega^{4} = 0.
	\end{equation}
	
	The master equation solution gives two values for the squared wave
	vector $k^{2}$ and hence two values for the squared phase velocity
	$v^{2} = \omega^{2}/k^{2}$.
	
	\subsection{Identical waveguide lines}
	
	In the case of identical lines, when
	$v_{01}^{2} = v_{02}^{2} = v_{0}^{2}$ and
	$\alpha_{1} = \alpha_{2} = \alpha$, the master equation solution is as
	follows:
	
	\begin{align}
		\label{eq:system6b}
		k^{2} &= \frac{\omega^{2}}{v_{0}^{2}}(1 \pm \alpha),\\
		v^{2} &= \frac{v_{0}^{2}}{(1 \pm \alpha)} \approx v_{0}^{2}(1 \mp \alpha),\\
		k_{1,2} &= \frac{\omega}{v_{0}}\left( 1 \pm \frac{\alpha}{2} \right)=k_0 \pm \Delta k,\\
		v_{1,2} &= \frac{v_{0}}{\left( 1 \pm \frac{\alpha}{2} \right)} \approx v_{0}\left( 1 \mp \frac{\alpha}{2} \right)=v_0\mp \Delta v,
		\label{eq:system6e}
	\end{align}
	where $k_0=\omega/v_0$, $\Delta k=\left({\alpha}/{2}\right)k_0$, and $\Delta v=\left({\alpha}/{2}\right)v_0$.
	
	Thus, the wave process in both lines consists of two waves having different phase velocities. The amplitude ratio can be easily obtained from the equations set \cref{eq:system5b,eq:system5e}:
	\begin{equation}
		\chi_{1,2} \equiv \left. \ \frac{B}{A} \right|_{k_{1,2}} = \frac{\alpha\omega^{2}}{\left( k^{2}v_{0}^{2} - \omega^{2} \right)} = \frac{\alpha}{(1 \pm \alpha) - 1} = \pm 1.
	\end{equation}
	And therefore, the wave process in the lines can be written as follows:
	\begin{align}
		U_{1} &= A_{1}\cos\left( \omega t - k_{1}x \right) + A_{2}\cos\left( \omega t - k_{2}x \right),\\
		U_{2} &= A_{1}\cos\left( \omega t - k_{1}x \right) - A_{2}\cos\left( \omega t - k_{2}x \right).
	\end{align}
	
	\begin{figure}[b!]
		\centering
		\includegraphics[width=8cm]{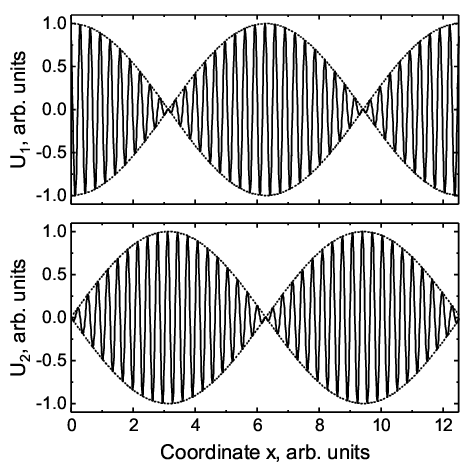}
		\caption{\label{fig:strong}The beating wave process in
			two coupled identical waveguide lines, when rf signal source $U = A\cos(\omega t)$ is connected to the input of the only first line.}
	\end{figure}
	
	When rf signal source $U = A\cos(\omega t)$ is applied to input of the
	only one of these two lines, \textit{e.g.} to the first line, it
	corresponds to the following simple boundary conditions:
	\begin{equation}
		A_{1} + A_{2} = A,\quad A_{1} - A_{2} = 0,
	\end{equation}
	resulting in amplitude values
	\begin{equation}
		A_{1} = - A_{2} = A/2.
	\end{equation}
	and hence in the following wave dynamics:
	\begin{multline}
		U_{1} = A \cdot \cos\left(\frac{k_{1} - k_{2}}{2}x \right)\cos\left( \omega t - \frac{{k_{1} + k}_{2}}{2}x \right) = \\
		= A \cdot cos(\alpha k_{0}x)\cos\left( \omega t - k_{0}x \right),
	\end{multline}
	\begin{multline}
		U_{2} = A \cdot \sin\left( \frac{k_{1} - k_{2}}{2}x \right)\sin\left( \omega t - \frac{{k_{1} + k}_{2}}{2}x \right) = \\ = A \cdot \sin\left( \alpha k_{0}x \right)\sin\left( \omega t - k_{0}x \right).
	\end{multline}
	This is a beating process corresponding to the periodic transfer of all wave energy from one waveguide line to the second line and back as shown in  \cref{fig:strong}. Any coupling between identical waveguide lines leads to the very strong interaction responsible for such maximal swing of the energy circulation, but spatial period of the beats
	\begin{equation}
		\label{eq:CycleProc}
		\Lambda=2\pi/\left(k_1-k_2\right)
	\end{equation}
	depends on the coupling factor value as follows:
	\begin{equation}
		\label{eq:LambdaIndentical}
		\Lambda=2\pi/\left(\alpha k_0\right).
	\end{equation}
	This means that $\Lambda$ is greater than the wave length by factor
	$\alpha^{-1}$.
	
	\subsection{Nonidentical waveguide lines}	
	
	To characterize interaction of two coupled waveguide lines quantitatively, in common with two coupled oscillatory circuits \cite{Migulin1983}, one can introduce degree of coupling of two waveguide lines with different phase velocities $v_{01}$ and $v_{02}$ as follows:
	\begin{equation}
		\label{eq:kappadef}
		\kappa = 2\sqrt{\alpha_1\alpha_2}\frac{v_{01}v_{02}}{\left|v_{02}^2 - v_{01}^2 \right|}.
	\end{equation}
	The degree of coupling has always finite value for nonidentical lines, but
	tends to infinity for identical lines without reference to the coupling
	factor $\alpha$ of the lines.
	
	In the case of nonidentical waveguide lines, the expressions for
	$k^2$ and $v^2$ have a much more complicated form as it follows
	from the master equation solution:
	\begin{equation}
		k^2 = \frac{\omega^2\left(v_{02}^2 + v_{01}^2\right)}{2v_{01}^2v_{02}^2} \pm \sqrt{\frac{\omega^4\left(v_{02}^2 + v_{01}^2\right)^2}{4v_{01}^4v_{02}^4} - \omega^4\frac{1 - \alpha_1\alpha_2}{v_{01}^2v_{02}^2}}.
	\end{equation}
	
	However, at weak interaction of the nonidentical lines, when
	\begin{equation}
		4\alpha_{1}\alpha_{2}\frac{v_{01}^2v_{02}^2}{\left| v_{02}^2 - v_{01}^{2} \right|^{2}} \ll 1,
	\end{equation}	
	the intricate expression for $k^{2}$ can be simplified as follows (assuming that
	$v_{02}^{2} > v_{01}^{2})$:
	\begin{equation}
		k_{1,2}^{2} = \frac{\omega^{2}}{v_{01,02}^2} \pm \frac{\alpha_1\alpha_2\omega^2}{\left(v_{02}^2 - v_{01}^2\right)},
	\end{equation}
	and hence
	\begin{equation}
		k_{1} = k_{01} + \Delta k_1,\qquad k_2 = k_{02}-\Delta k_2,
	\end{equation}
	\begin{equation}
		\label{eq:veloc}
		v_{1} = v_{01} - \Delta v_1,\qquad v_2 = v_{01} + \Delta v_2,
	\end{equation}	
	where
	\begin{equation}
		k_{01} = {\omega}/{v_{01}},\qquad k_{02} = {\omega}/{v_{02}},
	\end{equation}
	\begin{equation}
		\Delta k_{1} \simeq {\frac{\alpha_{1}\alpha_{2}v_{01}^{2}}{2\left( v_{02}^{2} - v_{01}^{2} \right)}}k_{01},\qquad	
		\Delta k_{2} \simeq {\frac{\alpha_{1}\alpha_{2}v_{02}^{2}}{2\left( v_{02}^{2} - v_{01}^{2} \right)}}k_{02},
	\end{equation}
	and
	\begin{equation}
		\Delta v_{1} = \frac{\alpha_{1}\alpha_{2}v_{01}^{2}}{2\left( v_{02}^{2} - v_{01}^{2} \right)}v_{01},\qquad
		\Delta v_{2} = \frac{\alpha_{1}\alpha_{2}v_{02}^{2}}{2\left( v_{02}^{2} - v_{01}^{2} \right)}v_{02}.
	\end{equation}	
	Then, the equation set \cref{eq:system5b,eq:system5e} gives the following amplitude ratios:
	\begin{align}
		\chi_{1} &\equiv \left.\frac{B}{A} \right|_{k_{1}} = - \frac{\alpha_{2}v_{01}^{2}}{\left( v_{02}^{2} - v_{01}^{2} \right)},\\
		\chi_{2} &\equiv \left.\frac{B}{A} \right|_{k_{2}} = \frac{\left( v_{02}^{2} - v_{01}^{2} \right)}{\alpha_{2}v_{01}^{2}}.
	\end{align}	
	In the case of low degree of coupling, coefficient $\left| \chi_{1} \right|$
	is small, while $\left| \chi_{2} \right|$ is oppositely high.
	
	\begin{figure}[t!]
		\centering
		\includegraphics[width=8cm]{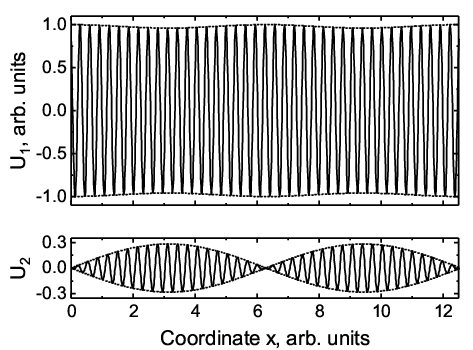}
		\caption{\label{fig:weak}The beating wave process in two coupled nonidentical waveguide lines at weak interaction of the lines ($\kappa\ll1$), when rf signal source $U = Acos(\omega t)$ is connected to input of the only first line.}
	\end{figure}
	
	When a rf signal source $U = Acos(\omega t)$ is applied to the input of the
	only first line, boundary conditions
	\begin{equation}
		A_{1} + A_{2} = A,\quad \chi_{1}A_{1} + \chi_{2}A_{2} = 0,\qquad\chi_1A_1+\chi_2A_2=0
	\end{equation}
	yield in the following amplitude values:
	\begin{equation}
		A_{1} = \frac{{\chi_{2}}_{}}{(\chi_{2} - \chi_{1})}A,\quad
		A_{2} = - \frac{{\chi_{1}}_{}}{(\chi_{2} - \chi_{1})}A.
	\end{equation}	
	Thus, the wave process in the coupled nonidentical lines can be written
	as follows:
	\begin{multline}
		U_{1} = A_{1}\cos\left( \omega t - k_{1}x \right) + A_{2}\cos\left( \omega t - k_{2}x \right) = \\ = D(x)\cos\left( \omega t - k_{1}x + \theta(x) \right),
	\end{multline}
	\begin{multline}
		\label{eq:U2}
		U_{2} = A_{1}\cos\left( \omega t - k_{1}x \right) - A_{2}\cos\left( \omega t - k_{2}x \right) = \\
		= 2A\frac{\left| \chi_{1} \right|\chi_{2}}{\left( \chi_{2} - \chi_{1} \right)}\sin\left( \frac{{(k}_{1} - k_{2})}{2}x \right)\times \\\times\sin\left( \omega t - \frac{{(k}_{1} + k_{2})}{2}x \right),
	\end{multline}
	where
	\begin{equation}
		\label{eq:Ampl49}
		D(x) = A_{1}\sqrt{1 + 2cos\left( \frac{k_1-k_2}{2}x \right)\frac{A_{2}}{A_{1}} + \left( \frac{A_{2}}{A_{1}} \right)^{2}},
	\end{equation}
	\begin{equation}
		\label{eq:phase50}
		\tan(\theta) = \frac{A_{2}\sin\left( \frac{k_1-k_2}{2}x \right)\ }{A_{1} + A_{2}\cos\left( \frac{k_1-k_2}{2}x \right)}
		\approx \frac{A_{2}}{A_{1}}\sin\left( \frac{k_1-k_2}{2}x \right).
	\end{equation}
	
	This is also a beating wave process, but with much less energy, which is transferred from the first line to the second one and back, as shown in \cref{fig:weak}. Therefore, the wave propagating along the first line has the only minor cyclic variation in amplitude \cref{eq:Ampl49} and phase \cref{eq:phase50}, while the wave penetrated into the second line represents low-amplitude beats accompanied with phase change by $\pi$ every beat cycle. Increase in the coupling factors $\alpha_1$, $\alpha_2$ decreases the spatial beat period:
	\begin{equation}
		\label{eq:LambdaNonIndentical}
		\Lambda = 2\pi/\left(k_{1} - k_{2}\right)= 2\pi/\left\lbrack\left(k_{01} - k_{02}\right) + \Delta k_1 + \Delta k_2 \right\rbrack,
	\end{equation}
	which is always less than $\Lambda_0 = 2\pi/\left(k_{01} - k_{02}\right)$.
	
	\section{Two-line JTWPA critical issues}
	
	In the development of JTWPAs based on using two artificial waveguide
	lines, one must take into account several critical issues arising from
	both the line coupling and the line discreteness.
	
	\subsection{Influence of coupling}
	
	In the two-line JTWPA designs, the inevitable mutual electrical or magnetic coupling between the waveguide lines causes a background beat process, producing the cyclic transfer of the traveling waves energy from one line to another and back. The nonlinear wave interaction required for the parametric amplifications has to coexist with this beat process. 
	
	In case of the flux-driven JTWPA mentioned above, the pump wave, propagating in the linear waveguide line, in reality will concurrently (i) modulate the inductances of the dc SQUID cells of the signal line and (ii) penetrate cyclically into this line with changing its phase by $\pi$ every beat cycle. The spatial beat cycle length and the maximum amplitude of the penetrated wave depend on a geometric average of the coupling factors of the lines and degree of their coupling $\kappa$, respectively
	
	As for the attractive idea to use the linear waveguide line as a pump energy source to balance (through the used coupling elements) depletion of the pump wave power in the other (signal) waveguide line and support the fixed pump wave amplitude in this line (when the pump and signal sources are connected separately to inputs of these linear and nonlinear waveguide lines, respectively), this cannot be properly realized since the only beating processes can occur in the two lines under mutual coupling. Besides, in force of the cyclic $\pi$-change in phase of the penetrated pump wave, the length of the amplifier line has to be restricted by the length $\Lambda$ of the spatial beat cycle given by \cref{eq:LambdaIndentical,eq:LambdaNonIndentical}. Moreover, the limitation on the pump wave amplitude, imposed by the critical current value of the used Josephson junctions, can lead to an even stronger restriction on the amplifier length especially at high degree of coupling of the lines. These facts impose restrictions on the attainable gain of the JTWPA. In addition, one should take into account the inevitable beat-like leakage of both the signal and idler waves into the other line and cyclic variations in their phases that further decreases the gain.
	
	\begin{figure}[b!]
		\centering
		\includegraphics[width=8.3cm]{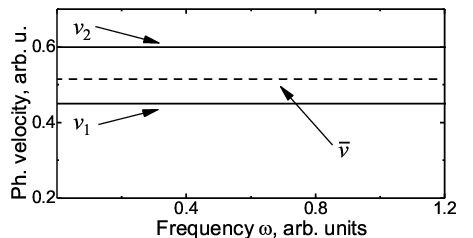}
		\caption{\label{fig:phasevelcont}Phase velocities $v_{1}$ and $v_{2}$ (solid lines) of the two eigen waves in the system of two coupled continuum waveguide lines at low degree of coupling. The dashed line shows the phase velocity $\overline{v}$ corresponding to the mean wave vector $\left(k_{1} + k_{2}\right)/{2}$}.
	\end{figure}
	
	The other significant drawbacks and restrictions follow from the inevitable phase mismatch between the pump wave and the signal and idler waves in the most interesting case of a low degree of coupling between the waveguide lines, when the only small part of the travelling wave energy is transferred cyclically between the lines. In the continuum
	approximation, the phase velocities \cref{eq:veloc} of the signal wave and the
	pumping wave, propagating in the first (nonlinear) line and in the
	second (linear) line respectively, are always different. In case of the
	flux driven JTWPA design, it makes impossible setting equal velocities
	of the of the signal wave and the inductance modulation process caused
	by the pump wave. Moreover, the phase velocity \(\overline{v}\) of the pump
	wave penetrated into the signal line (see \cref{eq:U2}) differs also from
	the signal velocity as shown in \cref{fig:phasevelcont}. However, these restrictions can be mitigated in the case of the discrete waveguide lines as shown below.

	\subsection{Influence of discreteness}
	
\begin{figure}[b!]
		\centering
		\includegraphics[width=9cm]{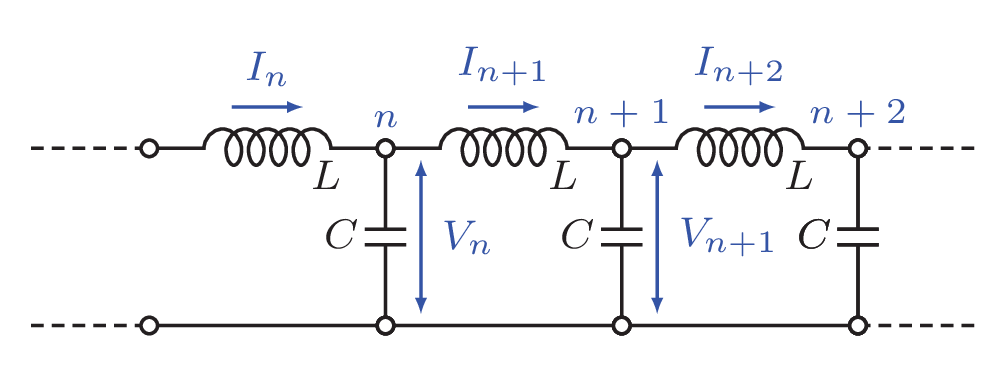}
		\caption{\label{fig:ArtifLC}An artificial discrete \textit{LC} line.}
	\end{figure}

	An artificial lossless discrete \textit{LC} waveguide lines (see \cref{fig:ArtifLC}) should be descried by the following discrete telegraph equations:
	\begin{align}
		I_{n} - I_{n + 1} &= i\omega CU_{n}, \label{eq:LCtI}\\
		U_{n} - U_{n + 1} &= i\omega LI_{n + 1}, \label{eq:LCtU}
	\end{align}
	where the propagating wave is considered as a harmonic wave. Let the cell size $a=x_{n+1}-x_n$ of the line be equal $a=1$, then
	both the wave voltage and current can be written as follows \cite{Kogan2024}:
	\begin{align}
		U_n& =U_0e^{i\omega t}e^{-ink}, \label{eq:Un}\\
		I_n&=I_0e^{i\omega t}e^{-ink}, \label{eq:In}
	\end{align}
	where $U_0$ and $I_0$ are the initial complex amplitudes of
	voltage and current, and $k$ is the wave number. In this case, the
	discrete telegraph equations \cref{eq:LCtI,eq:LCtU} result in the following dispersion expression:
	\begin{equation}
		\sin^{2}\left(k/2\right) = \frac{\omega^{2}}{\omega_{cut}^{2}}
	\end{equation}
	and the phase velocity
	\begin{equation}
		v = \frac{\omega}{k} = \frac{\omega}{2\arcsin\left( {\omega}/{\omega_{\text{cut}}} \right)}=\frac{\omega}{2\theta},\label{eq:phasevel}
	\end{equation}
	where $\omega_{\text{cut}}$ is the upper cut-off frequency of the discrete line:
	\begin{equation}
		\omega_{\text{cut}} = \frac{2}{\sqrt{LC}}\label{eq:wcut}
	\end{equation}
	and
	\begin{equation}
		\theta = \arcsin\left( \frac{\omega}{\omega_{cut}} \right).\label{eq:Theta}
	\end{equation}
	
	Moreover, such a discrete line generally has a complex wave impedance
	with the imaginary and real parts both depending on frequency \cite{Nikolaeva2023}. In contrast to a continuous waveguide line, the wave impedance $Z_0$ of a discrete \textit{LC} line composed of lumped $L$ and $C$ elements (as shown in \cref{fig:JTWPA_idea}) cannot be defined and
	measured as an exactly local ratio of the wave voltage to the wave
	current, since the voltage values are defined only at the line nodes,
	while the wave current values are measured in the lumped $L$
	elements between the nodes. Therefore, the discrete telegraph equations
	\cref{eq:LCtI,eq:LCtU} with \cref{eq:Un,eq:In} give the following wave impedance $Z_{0}$ (see also the other derivation of $Z_{0}$ as a repetitive impedance of the lumped cell array in  \cite{Nikolaeva2023}):
	\begin{equation}
		Z_0= \sqrt{L/C}\cdot e^{\pm i\theta}=\pm i\omega L/2+\sqrt{L/C-\left(\omega L\right)^2/4},
	\end{equation}
	were the two signs ``$+$'' and ``$-$'' correspond to the ratios $V_n⁄I_n$  and $V_n⁄I_{n+1}$, respectively, and \(\theta\) is defined by \cref{eq:Theta}. This formula describes also the input impedance of the discrete line when it starts either with $L$ element (sign ``$+$'') or with initial $C$ element (sign ``$-$''). It can be seen that the imaginary part of $Z_0$ increases with frequency, while the real part of $Z_0$ decreases from the value $\sqrt{L/C}$ at low frequency down to zero at the cut-off frequency. When the discrete line is composed of symmetric T-cells ``$L/2$--$C$--$L/2$'' and hence the line starts with initial $L/2$ element, its input impedance becomes completely real
	but again frequency-dependent:
	\begin{equation}
		Z_0=\sqrt{L/C-\left(\omega L\right)^2/4}.
	\end{equation}
	This dependence on frequency substantially obstructs perfect matching of
	the line in a wide frequency band. For example, when the discrete line
	is connected to the real-valued impedance $\rho = \sqrt{L/C}$, it
	results in the following imaginary and frequency-dependent reflection
	factor:
	\begin{equation}
		\Gamma = \left(\rho - Z_{0}\right)/\left(\rho + Z_0\right) = i \cdot \tan\left(\theta/2\right).
	\end{equation}
	
	The existence of the upper cut-off frequency \cref{eq:wcut} makes possible keeping	both the higher harmonics of the pump signal and the unwanted
	intermodulation components out of the frequency band. At the same time,
	the frequency-dependent phase velocity \cref{eq:phasevel} in the frequency band does not allow achieving good phase matching between the pumping, signal and	idler waves, needed to attain a good amplification. Therefore, an
	additional dispersion engineering techniques (\textit{e.g.} see \cite{Remm2023,Perelshtein2022}) has to be used to improve the wave phase-matching.
	
	In a one-line JTWPA, the phase mismatching can be mitigated with using the 4-wave-mixing (4WM) mode \cref{eq:4W,eq:4Wk}. In the mode, all the frequencies
	$\omega_{p}$, $\omega_{s}$, $\omega_{i}$ are located near each
	other (contrary to 3WM mode) and hence can be set in the vicinity of
	$\omega_{cut}/3$, where the wave velocity slowly changes with
	frequency. Besides, both the harmonics of pump wave and the unwanted
	intermodulation components produced by the cubic nonlinearity appear
	above the cut-off frequency.
	
	\begin{figure}[t!]
		\centering
		\includegraphics[width=8.3cm]{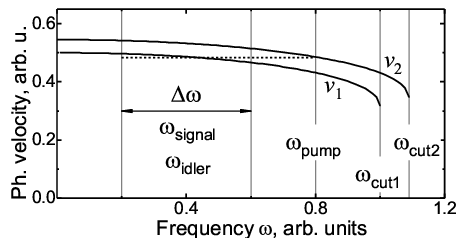}
		\caption{\label{fig:PhaseVel}The phase velocities $v_1$ and $v_2$ (corresponding to the eigen wave vectors $k_1$ and $k_2$, respectively) of the waves propagating of the discrete coupled nonlinear (Josephson)and linear lines with different cut-off frequencies at low degree of coupling of the used lines, as well as the possible frequency scheme corresponding to 3WM
			mode, which can be realized in the flux-driven JTWPA. Here
			$\Delta\omega$ is the frequency range, where the phase velocities of the signal and idler waves in the Josephson line (lower curve) are close to the phase velocity of the pump wave propagating in the linear line (upper curve). The cut-off frequencies of the coupled nonlinear (Josephson) and
			linear lines are $1.0\text{ au}$ and $1.09\text{ au}$, respectively.}
	\end{figure}
	
	In the case of a hypothetical two-line JTWPA, one can recommend to use the 3-wave-mixing (3WM) mode \cref{eq:3W,eq:3Wk} instead as a better mismatching trade-off. \Cref{fig:PhaseVel} shows the possible frequency scheme corresponding to 3WM mode, which can be realized in the flux-driven JTWPA. In the scheme, $\Delta\omega$ is the frequency range, where the phase velocities of the signal and idler waves in the Josephson (signal) line (curve $v_1$) are close to the phase velocity of the pump wave propagating in the linear line (curve $v_2$) and modulating inductances of the dc SQUID cells of the signal line. These velocities correspond to the eigen wave vectors $k_1$ and $k_2$ for the system of two coupled discrete waveguide lines with different cut-off frequencies at low degree of coupling of the lines. Of course, some small part of the pump wave power will also penetrate into the signal line and decrease available linear dynamic range of the amplifier. As shown in \cite{Nikolaeva2023}, an additional increase in the reference dynamic range can be achieved with substituting dc SQUIDs for bi-SQUIDs.
	
\begin{figure}
		\centering
		\includegraphics[width=8.3cm]{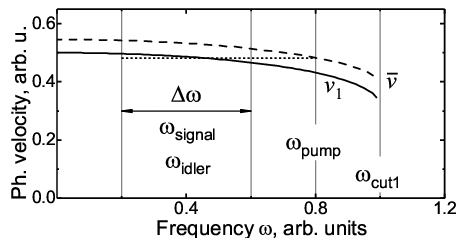}
		\caption{\label{fig:PhaseVelFD}The phase velocities $v_1$ (corresponding to the eigenwave vector $k_1$) of the wave propagating in the signal Josephson
			line (solid curve) and the phase velocity $\overline{v}$ (corresponding to the wave vector $\overline{k} = \left(k_1 + k_2 \right)/2$) of the pump wave
			penetrated into this signal line (dashed curve) at low degree of coupling of the used lines, as well as the possible frequency scheme corresponding to 3WM mode in the signal Josephson line. 
			Here $\Delta\omega$ is the frequency range, where the phase
			velocities of the signal and idler waves are close to the phase velocity
			$\overline{v}$ of the pump wave penetrated into the signal line. The
			cut-off frequencies of the coupled nonlinear (Josephson) and linear
			lines are $1.0\text{ au}$ and $1.2\text{ au}$, respectively.}
	\end{figure}

	If one considers the two-line design where the pumping wave in the nonlinear Josephson line is supplied through the used coupling elements by the high-power wave propagating in the other waveguide line which is linear, the 3WM mode can be suggested in accordance with the frequency diagram shown in \cref{fig:PhaseVelFD}. This diagram is similar to the one presented in \cref{fig:PhaseVel}, where the phase velocity of the pump wave $v_2$ in the linear line is replaced by the phase velocity $\overline{v}$ (dashed line) corresponding to the mean wave vector $(k1 + k2)/2$. This is the phase velocity of the pump wave penetrated into the nonlinear Josephson line as follows from \cref{eq:U2}. Unfortunately, a nontrivial trade-off has to be determined for both the parameters of the waveguide lines (resulting in the cut-off frequencies) and their mutual coupling in order to set the compromised values of (i) the phase velocity $\overline{v}$ of the penetrated pump wave, (ii) the amplitude $A_p$ of the wave (see \cref{eq:U2}), and (iii) the spatial beat cycle length $\Lambda$ given by expression \cref{eq:LambdaNonIndentical}.
	
	\section{Conclusion}
	
	In summary, there are several critical obstacles to the two-line designs of Josephson traveling wave parametric amplifiers (JTWPA) such as the flux-driven JTWPA \cite{Zorin2019, Nikolaeva2023} and the other ones, where an additional linear waveguide line is supposed to be used as a pump wave energy source for the nonlinear Josephson line in order to balance (through coupling elements) the pump wave power depletion and hence to support the fixed pump wave amplitude in this waveguide line. The obstacles result from both (i) the cyclic transfer of wave energy from one waveguide line to the other one and back and (ii) the phase mismatch between the pump, signal and idler waves following from the line coupling and the line discreteness. In particular, in the case of flux-driven JTWPA, the required traveling-wave-patterned modulation of the SQUID-cell inductances of the Josephson signal line under the influence of a high-power pump wave, will be inevitably accompanied with an unwanted cyclic penetration of the pump wave to the nonlinear Josephson line. Besides, the cyclic phase variation and the cyclic power leakage of the signal and idler waves to the pump line will also interfere with the amplification process.

	\section*{Conflict of Interest Statement}
	\noindent The authors declare no conflict of interest. The funder had no role in any part of the manuscript preparation process.

	\bibliographystyle{IEEEtran}
	\bibliography{Kornev_TwoLineJTWPA}

\end{document}